\documentclass[pra, aps,twocolumn,superscriptaddress,longbibliography,floatfix,notitlepage]
{revtex4-2}

\usepackage{textcomp}
\usepackage[utf8x]{inputenc}
\usepackage[english]{babel}
\usepackage[T1]{fontenc}
\usepackage{lmodern}
\usepackage{amsmath,amsfonts,amssymb,amsthm,bm,times,dcolumn}
\usepackage{mwe}
\usepackage{microtype}
\usepackage{braket}
 
\usepackage{gensymb} 
\usepackage{physics}
\usepackage[colorlinks={true}, citecolor={blue}, filecolor={blue}, linkcolor={blue}, urlcolor={blue}]{hyperref}
\usepackage{graphicx,color}
\usepackage[caption=false]{subfig}

\begin{document}
\pdfoutput=1
\title{A Computational and Experimental Analysis of Higher Order Modes in a Strongly Focusing Optical Cavity}

\author{Mehmet Öncü}
    \email[Electronic address:]{mehmet.oncu@ozu.edu.tr}
    \affiliation{Özyeğin University, Electrical and Electronics Engineering Department, Istanbul, 34794, Turkey}
\author{Mohsen Izadyari}
    \email[Electronic address:]{mizadyari18@ku.edu.tr}
    \affiliation{Koç University, Physics Department, Istanbul, 34450, Turkey}
\author{Özgür E. Müstecaplıoğlu}
    \email[Electronic address:]{omustecap@ku.edu.tr}
    \affiliation{Koç University, Physics Department, Istanbul, 34450, Turkey}
    \affiliation{TÜBİTAK Research Institute for Fundamental Sciences, 41470 Gebze, Turkey}
\author{Kadir Durak}
    \email[Electronic address:]{kadir.durak@ozyegin.edu.tr}
    \affiliation{Özyeğin University, Electrical and Electronics Engineering Department, Istanbul, 34794, Turkey}

\begin{abstract}
Optical cavities operating in the near-concentric regime are the fundamental tools to perform high precision experiments like cavity QED applications. A strong focusing regime unfortunately is prone to excite higher-order modes. Higher-order mode excitation is challenging to avoid for the realistic strong focusing cavities, and if these modes are closely spaced, overall cavity linewidth gets significantly broadened. In this study, a computational method alongside the experiment is provided for the optical mode decomposition into cavity eigenmodes with justified approximations. It is shown that it is possible to recreate the intensity and spectral profile of the cavity transmission, with the provided model. As a result, a more complete treatment of the realistic near-concentric cavities can be done.
\end{abstract}

\maketitle

\section{\label{sec:intro} Introduction}
 Optical cavities are at the heart of photonics applications, including profound quantum and non-linear effects, high optical power, and precise control of atoms and molecules ~\cite{dutra2005cavity, thompson1992observation, hood1998real, puppe2004single, reiserer2015cavity, brennecke2007cavity, clifford1998high, nguyen2017single, nguyen2018operating}. To acquire strong atom-photon coupling, high precision, and sensitivity demanded by such applications, detailed knowledge of the optical mode structure in the cavity is needed ~\cite{bond2011higher, utama2021coupling}. In practice, higher-order modes can also be excited in the strong coupling regime, together with the target optical mode ~\cite{durak2014diffraction, kleckner2010diffraction}.
 \par To model modern cavity QED experiments in strong coupling conditions, it is necessary to determine the coupling efficiency of higher-order modes. Moreover, each higher-order mode is subject to different diffraction losses and mode matching efficiencies ~\cite{benedikter2015transverse}. Hence, diffraction losses and reflectivity of the resonator mirrors should be considered for a realistic description of the mode structure of the cavity. A well-defined single mode for an optical cavity requires that the radius of curvatures of the resonator mirrors, placed at the nodes of the standing wave pattern, match that of the wavefront of the input optical mode. If the beam’s wavefront is not aligned perfectly with the radius of curvature of the resonator mirror, mode decomposition structure changes. In addition, the aberrations caused by the cavity mirror or the mode-matching optics perturb the system from the ideal situation and lead to excitation of the higher-order modes. This results in the cavity linewidth broadening of the cavity transmission, if the higher-order modes cannot be resolved ~\cite{hood2001characterization, jaffe2021aberrated}. If the higher-order modes are closely spaced, broadening makes the resolution of the fundamental mode impossible. Hence, determination of the detailed mode structure and the linewidth broadening due to higher-order modes in realistic models of optical cavities, which is our objective here, is significant for the efficient implementation of photonics applications. 

We construct an optical cavity experimentally in the strong focusing regime and apply a computational approach to calculate the cavity's mode structure and the linewidth broadening due to the higher-order mode excitations observed in the experiment. Our approach allows for the determination of the change in the overall cavity linewidth due to the different minimum beam waists, reflectivities, and mirror apertures, which can be particularly significant for applications in the strongly focused optical cavity regime.  The model offered in this paper allows us to study different parameters that are vital to the understanding of the near-concentric cavities. Furthermore, it can also be used to simulate all types of optical cavities with spherical mirrors for arbitrary incident optical mode with various phases, and it is the generalized version of the work given by Ref.~\cite{durak2014diffraction} . In this work, the effect of each higher order mode is analyzed, rather than their combined behavior which can be found in Ref.~\cite{durak2014diffraction, kleckner2010diffraction}.  Effects of different parameters like coupling efficiencies, diffraction losses, reflectivities, or beam waist sizes can be understood thoroughly with the help of the mode decomposition analysis presented in this work. Arbitrary shape optical modes can be studied further with the given model by changing the phase parameter in the incident mode definition. Investigating near-concentric cavity parameters in terms of each mode contributes to the complete understanding of these devices.

The rest of the paper is organized as follows. Sec.~\ref{sec:physical system} describes our experimental optical cavity setup. Sec.~\ref{sec:methods} introduces the computational methods we use to assess the significance of higher-order modes in the optical cavity. Sec.~\ref{sec:results and discussion} presents the experimental and computational results. We conclude in Sec.~\ref{sec:conclusion}. 

\section{\label{sec:physical system} Physical System}  

\begin{figure}
    \centering
    \includegraphics[width=\columnwidth]{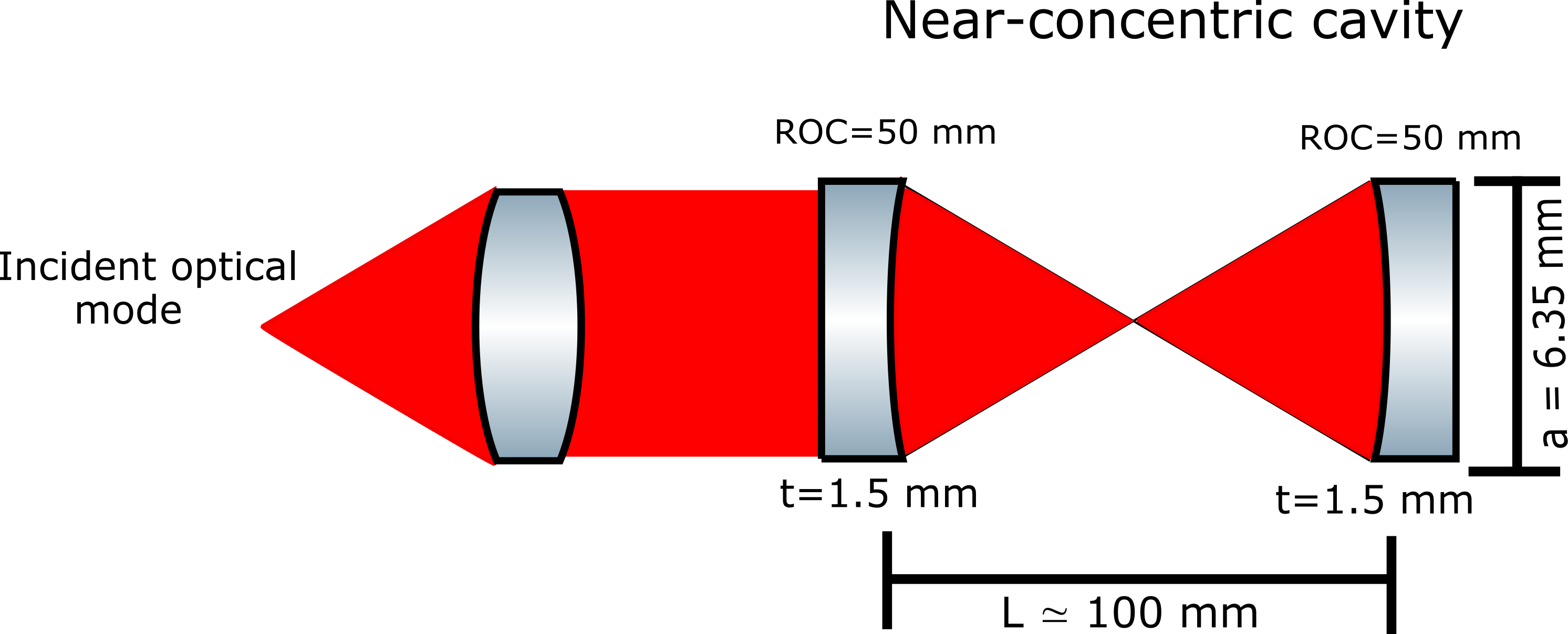}
    \caption{Basic diagram of the cavity used for simulation and experimental study. Here, radius of curvature of the mirrors (ROC) are 50 mm, thickness is $t=1.5$ mm, aperture size is $a=6.35$ mm and cavity length is $L=100$ mm.}
    \label{Setup diagram}
\end{figure}

Physical system consists of two spherical plano-concave mirrors whose geometrical and optical features are exactly the same.  Mirror sizes (a), radius of curvatures (ROC) and thicknesses (t) are taken as 6.35 mm, 50 mm and 1.5 mm, respectively. For the both simulation and the experiment, 780 nm light is used. Reflectivities are 0.9798 for the both mirrors. In the study, we use near-concentric cavity whose cavity length is nearly $L=100$ mm. Physical system is studied experimentally and computationally. Simple diagram for the experiment is seen in Fig.~\ref{Setup diagram}.

\section{\label{sec:methods} Methods}

Optical cavity modes can be described by different sets of polynomials. One of the widely used modes are Laguerre-Gaussian modes $\text{LG}_{lp}$ which can be written as the following ~\cite{siegman86},
\begin{equation}
\begin{gathered}
         \Psi_{lp}(r,\phi,z)= C_{lp} \frac{w_0}{w(z)} \frac{r\sqrt{2}}{w(z)}^{|l|} \text{exp} \Big( \frac{-r^2}{w(z)^2} \Big)  L_p^{|l|} \Big( \frac{2r^2}{w(z)^2} \Big) \\
         \times \text{exp} \Big(\frac{ikr^2}{2R(z)} \Big) \text{exp}\big(il\phi\big) \text{exp}\big([-i(2p+|l|+1)\xi(z)]\big)
         \label{LG}
\end{gathered}
\end{equation}
where $L_p^{|l|}$ are generalized Laguerre polynomials, $w(z)$ is mode waist at the position z, p is the radial index, l is the azimuthal index, R(z) is the radius of curvature of the wavefront at the position z, $\xi(z)$ is the Gouy phase which can be written in terms of the Rayleigh length, and $\text C_{lp}$ are the normalization coefficients. We consider a cylindrically symmetric optical cavity, described by $\text LG_{lp}$ modes and driven by an optical mode taken to be in the form of  $\text LG_{00}$, which can be written as ~\cite{durak2014diffraction, kleckner2010diffraction},
\begin{equation}
    \Phi(r,\phi,z)=C_{00}\frac{w_0}{w(z)} \text{exp} \Big(\frac{-r^2}{w(z)^2} \Big)  \text{exp}\Big(\frac{ikr^2}{2R(z)}\Big) \text{exp}\big(-i\xi(z)\big) .
    \label{OM}
\end{equation}

In the model, optical mode wavefront is taken as spherical, whose radius of curvature is exactly equal to the radius of curvature of the cavity mirrors. Simulation parameters are the same as those used in the experimental study, described in Sec.~\ref{sec:physical system}. In order to find the beam waist at the mirror and the cavity length, initially half cavity length is calculated for each minimum beam waist value by the following equation,
\begin{equation}
R(z)=z\Big( 1 + \Big( \frac{\pi w_0^2 n}{\lambda z} \Big)^2 \Big)   
\label{ROC}
\end{equation}
by finding the z value whose radius of curvature is exactly equal to the radius of curvature of the mirror. This z value gives the position of the mirror, which is substituted in the Eq.~(\ref{beam waist}) to find the beam waist at the mirror,
\begin{equation}
    w(z)=w_0 \Big( 1 + \Big( \frac{z}{z_0} \Big)^2 \Big)^{1/2}.
    \label{beam waist}
\end{equation}

It is assumed that the input beam has a perfect spherical wavefront without any azimuthal deviation, and a phase difference is introduced in the radial direction. We employ the paraxial approximation and calculated the phase of the optical mode by the standard ray-tracing methods ~\cite{durak2014diffraction, kleckner2010diffraction}. Coupling strength of the optical drive to a cavity mode $\text LG_{lp}$ is characterized by the mode overlap function,

\begin{equation}
    \gamma_{lp}=|\Psi_{lp}(r,\phi,z)^*\Phi(r,\phi,z)|^2  .
    \label{coupling eqn}
\end{equation}

\begin{figure}
    \includegraphics[width=2.55in]{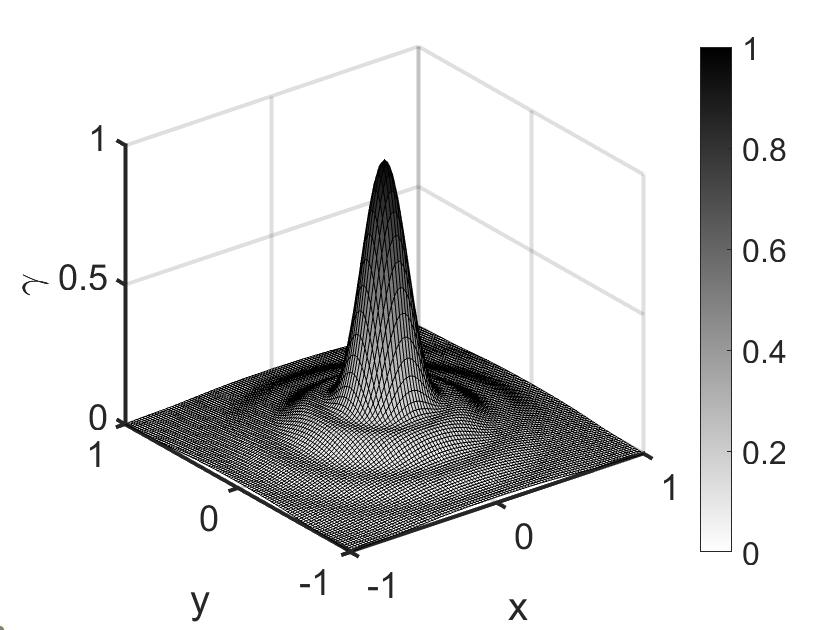}
    \caption{ Spatial profiles of the overlap function for the input $\text LG_{00}$ optical mode and the all cavity eigenmodes. $\gamma$ is the transmission inside the cavity, which is plotted spatially having x and y as the spatial axes.  Minimum beam waist is taken as $w_0=7.3\,\mu$m.}
    \label{Spatial and density distribution of optical mode coupled to a cavity eigenmodes}
\end{figure}

\begin{figure}
    \subfloat[]{\includegraphics[width=2in]{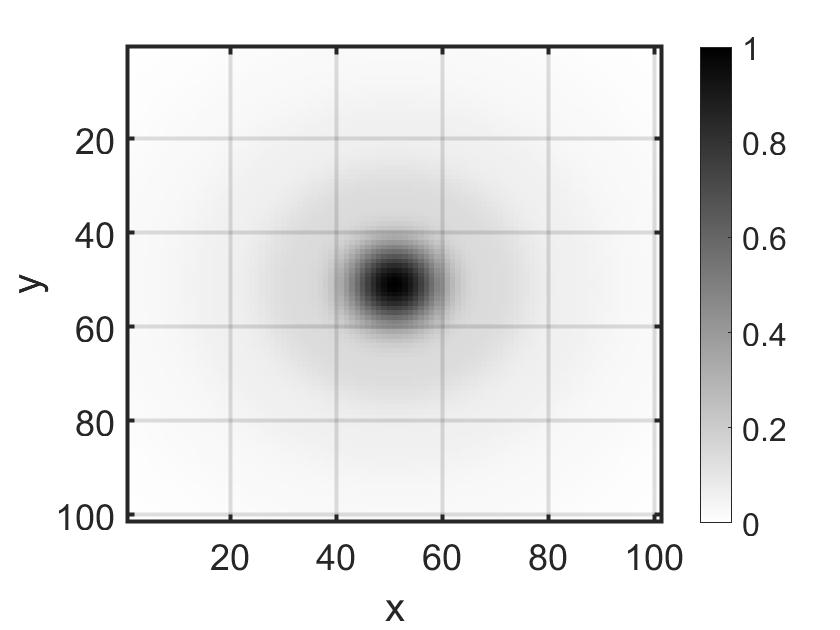}} 
    \subfloat[]{\includegraphics[width=1.4in]{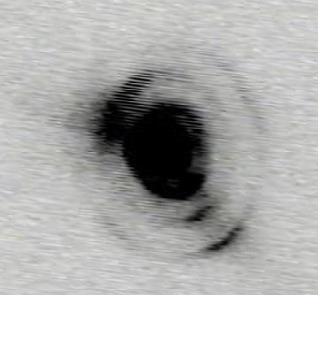}} 
    \caption{Heat map and the camera photograph showing the cavity field distribution. Camera photo is inverted to show the effect clearly.}
    \label{heat map and camera photo}
\end{figure}

It is an intuitive measure of the injected power distribution inside the cavity resonator and $\bar{\gamma_{lp}}$ is the normalized integral version of the Eq.~(\ref{coupling eqn}). Spatial profile $\gamma_{lp}$ of the cavity is given in Fig.~\ref{Spatial and density distribution of optical mode coupled to a cavity eigenmodes}. Spatial profile of the overlap function for the input mode and the set of cavity modes is seen in Fig.~\ref{Spatial and density distribution of optical mode coupled to a cavity eigenmodes}. The total overlap function is calculated for $\text l=0$ and $\text p=0,1,2,...50$ and their combined effect is shown in Fig.~\ref{Spatial and density distribution of optical mode coupled to a cavity eigenmodes}. The total overlap function changes to weighted summation over the modes in calculating optical properties, such as transmission of the cavity and power distribution inside the cavity. Corresponding heat map and the camera picture obtained from the experiment is shown in Fig.~\ref{heat map and camera photo}. As can be seen from the graph, heat map and the camera picture have a good matching profile.  

In order to calculate the cavity transmission, the amount of power left in the cavity after one round trip should be taken into account and it is given by $\alpha_{lp}$ which are calculated as ~\cite{durak2014diffraction},

\begin{equation}
   \alpha_{lp}= \abs{\frac{\int_0^a \int_0^{2\pi} |\Psi_{lp}|^2 r dr d\phi}{ \int_0^{\infty} \int_0^{2\pi} |\Psi_{lp}|^2 r dr d\phi  } }^2 
   \label{diff loss eqn}
\end{equation}
where a is the mirror aperture. The coefficient of finesse $\text F_{lp}$ is defined ~\cite{yariv2007photonics},

\begin{equation}
    F_{lp}=\frac{4 \rho_{lp}}{(1-\rho_{lp})^2}
    \label{F constant}
\end{equation}
where $\rho_{lp}$ is the reflection coefficient, which should be calculated separately for each cavity eigenmode by taking the mirror reflectivities and diffraction losses into account using,

\begin{equation}
    \rho_{lp}=R_m^2 \Big(\int_0^a \int_0^{2\pi} |\Psi_{lp}(r,\phi,z)|^2 r dr d\phi \Big)^2 .
    \label{rho eqn}
\end{equation}
Transmission of each higher order mode can be written as $ \text T_{lp}$. While calculating the power inside the cavity and the power
in each higher order mode, following equation is used ~\cite{lipson2010optical}, 
\begin{equation}
    T_{lp}=\alpha_{lp} \gamma_{lp} \frac{1}{1+F\sine^2{\delta/2}}
    \label{transmission}
\end{equation}
where $\delta$ is the phase difference between the interfering wavefronts which is in this case equal to $\delta=2 \pi \omega L / c$. With the coefficient of finesse $\text F_{lp}$, finesse $\mathcal{F}_{lp}$ can be written as ~\cite{yariv2007photonics},  
\begin{equation}
    \mathcal{F}_{lp} = \frac{\pi}{2} \sqrt{F_{lp}}
    \label{finesse}
\end{equation}

\section{\label{sec:results and discussion} Results and Discussion}
\begin{figure}
    \centering
    \includegraphics[width=2.5in]{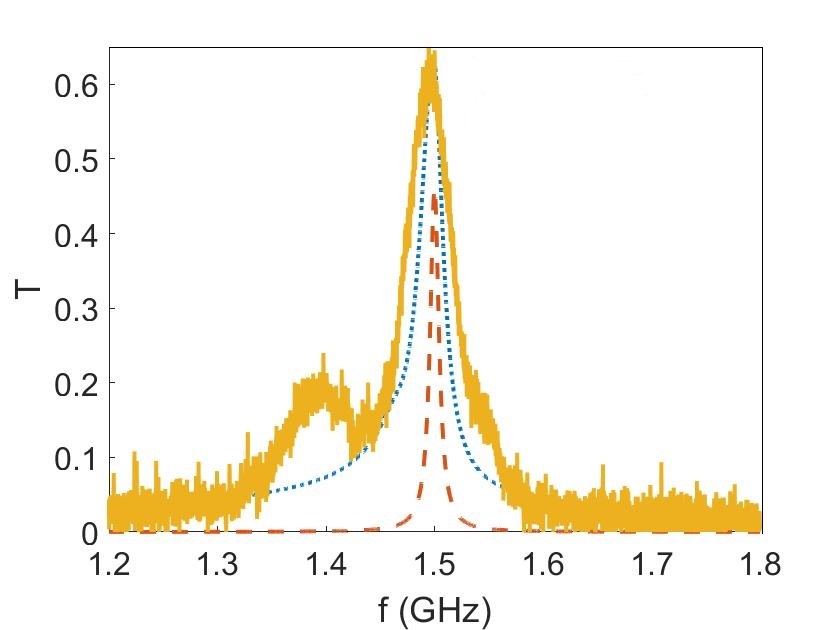}
    \caption{Cavity linewidth broadening effect shown experimentally and computationally with minimum beam waist of $w_0=7.3$ $\mu$m. T is the cavity transmission plotted for frequency (f) values. Solid yellow curve shows the experimental transmission, whereas dotted blue and dashed red curves are obtained computationally. Dashed red curve is for the fundamental cavity eigenmode $\text LG_{00}$ and dotted blue curve is for the overall cavity field calculated for the first 50 cavity eigenmodes.}
    \label{Experimental data}
\end{figure}

\begin{figure}
\subfloat[]{\includegraphics[width=1.8in]{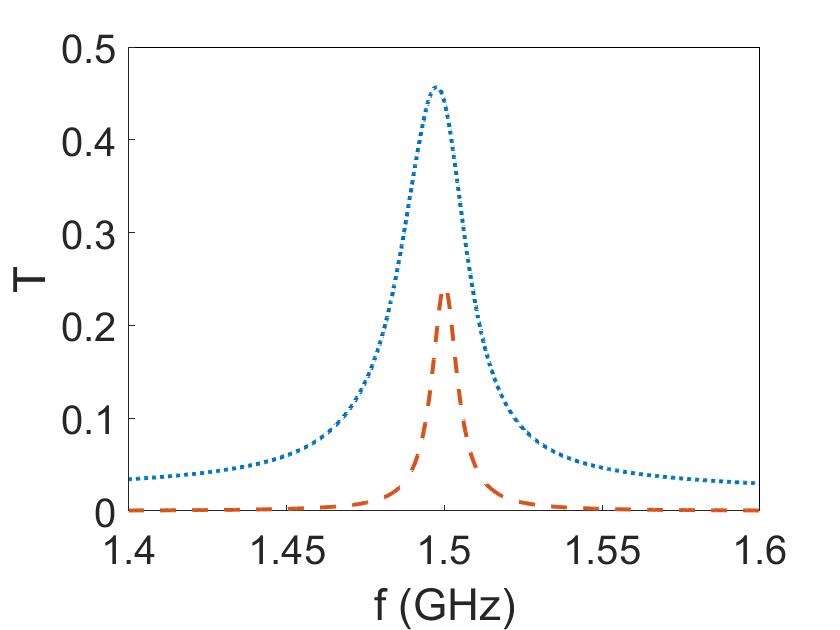}} 
\subfloat[]{\includegraphics[width=1.8in]{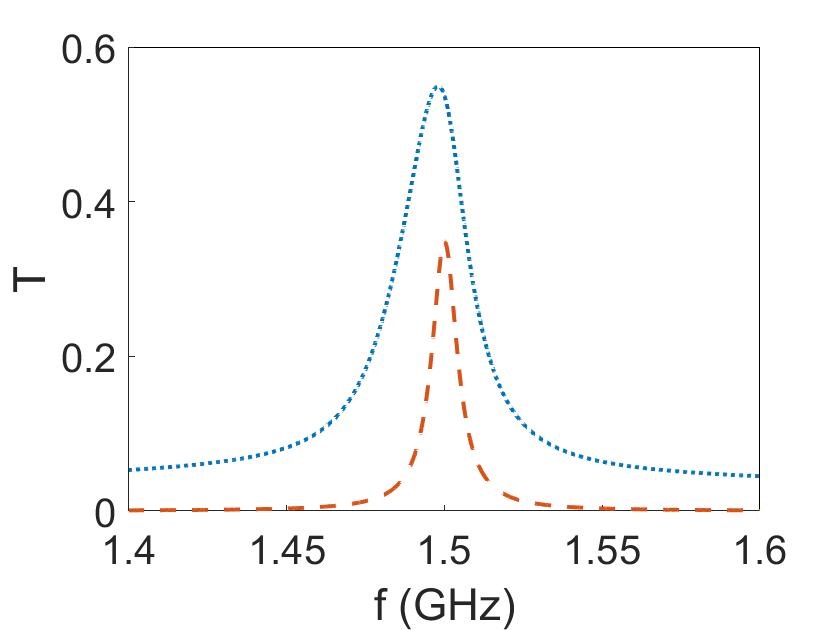}}\\
\subfloat[]{\includegraphics[width=1.8in]{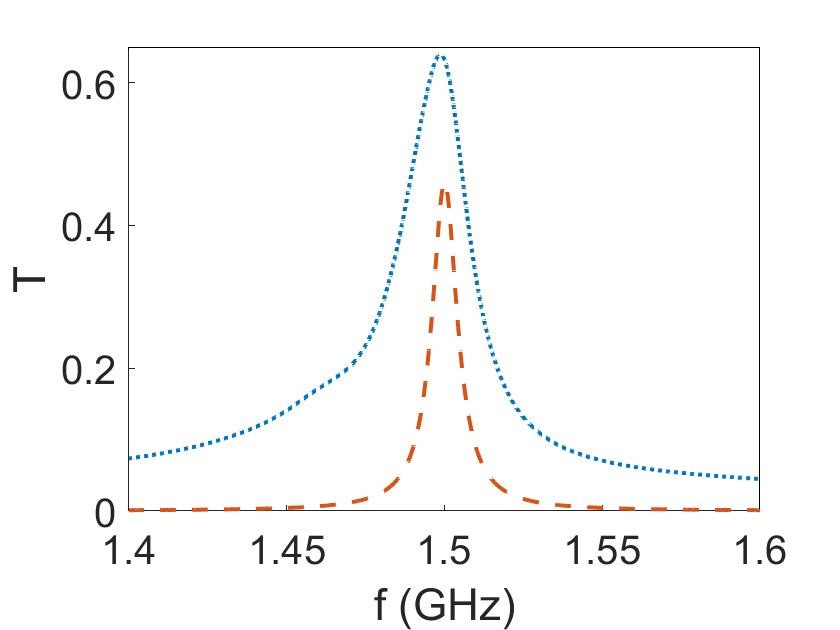}}
\subfloat[]{\includegraphics[width=1.8in]{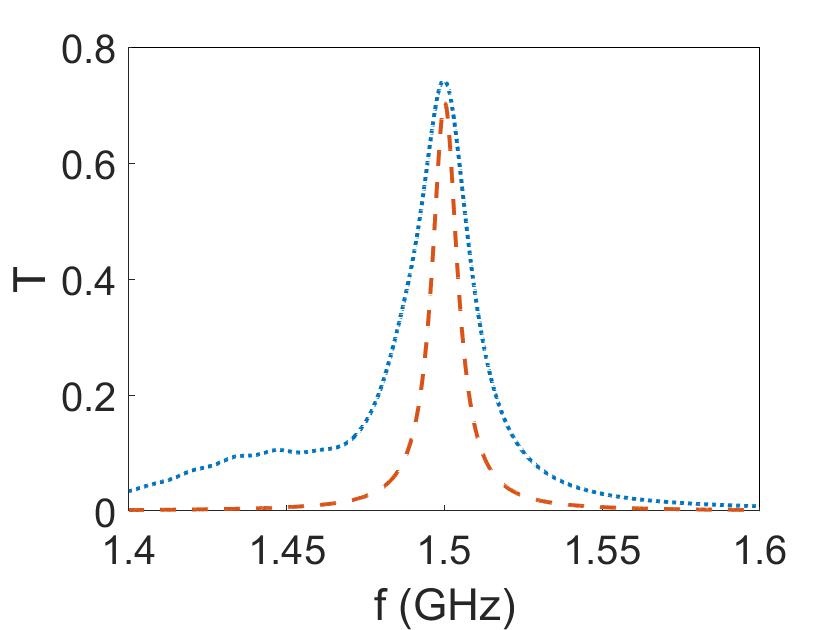}} 
\caption{Total transmission plotted against the frequency (f). The dashed red curve shows the fundamental mode whereas the dotted blue curve shows the effect of the linewidth broadening after contribution from each cavity eigenmode is calculated. Figures a) b) c) and d) demonstrate the cases where minimum beam waists are 5.7, 6.5, 7.3, and 9.3 $\mu$m, respectively. Total cavity linewidth values are $\kappa=26.324, 27.868, 28.002$ and $22.175$ MHz, respectively. }
\label{Linewidth broadening}
\end{figure}

Cavity in the near-concentric regime is experimentally realized as described in the Sec.~\ref{sec:physical system}, whose transmission graph is given in Fig.~\ref{Experimental data}. On the same graph, fundamental cavity eigenmode transmission and the overall field transmission can be seen. From the experimental data, shown with the yellow solid line, it is seen that the overall field does not only include the fundamental mode, but it is broadened due to the higher order mode excitations. Fig.~\ref{heat map and camera photo} and Fig.~\ref{Experimental data} prove that the model matches with the experimental scenario and similar patterns can be found in Ref.~\cite{durak2014diffraction} since the same type of optical cavities are used.

\begin{figure}
\subfloat[]{\includegraphics[width=1.8in]{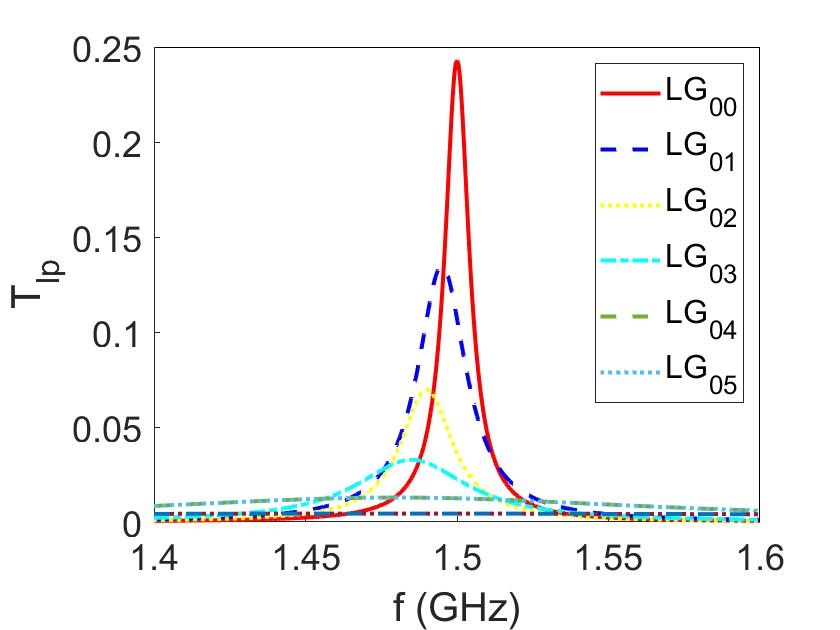}} 
\subfloat[]{\includegraphics[width=1.8in]{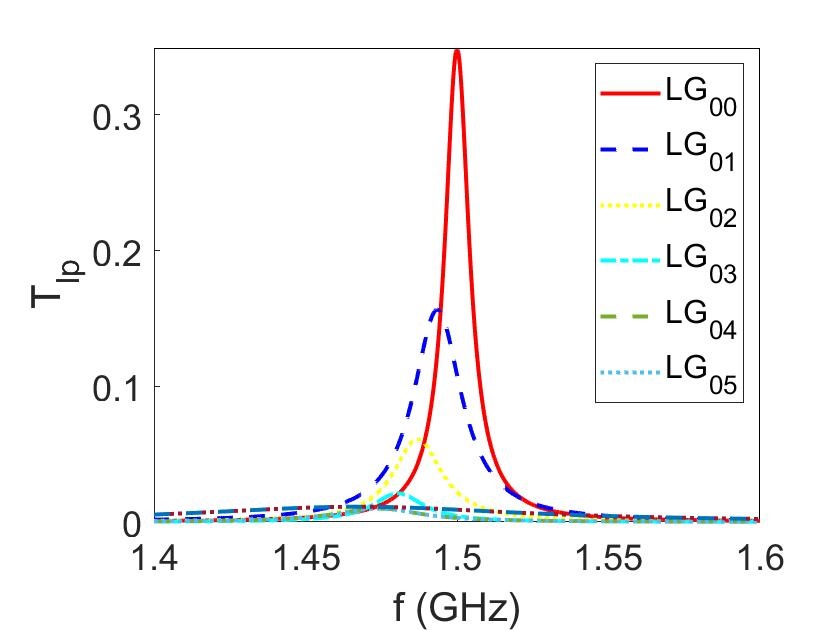}}\\
\subfloat[]{\includegraphics[width=1.8in]{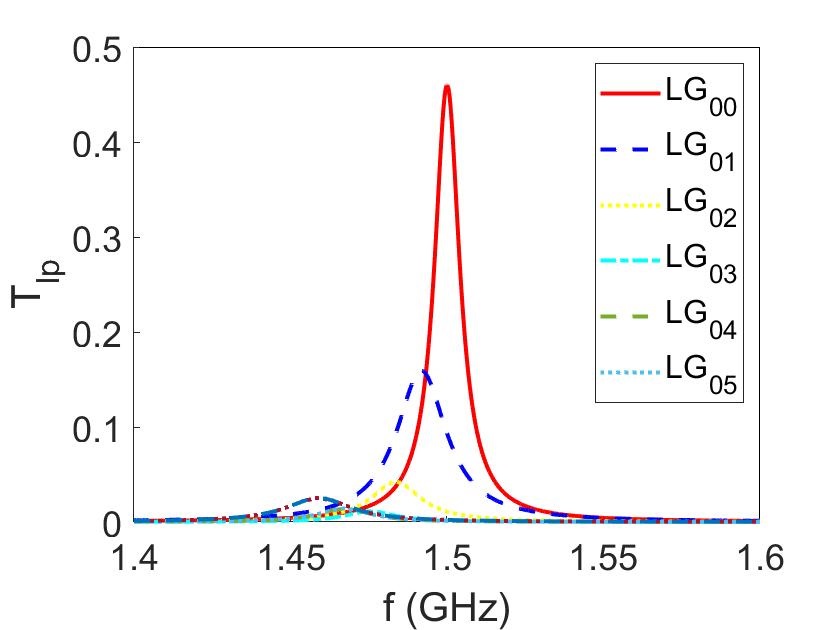}}
\subfloat[]{\includegraphics[width=1.8in]{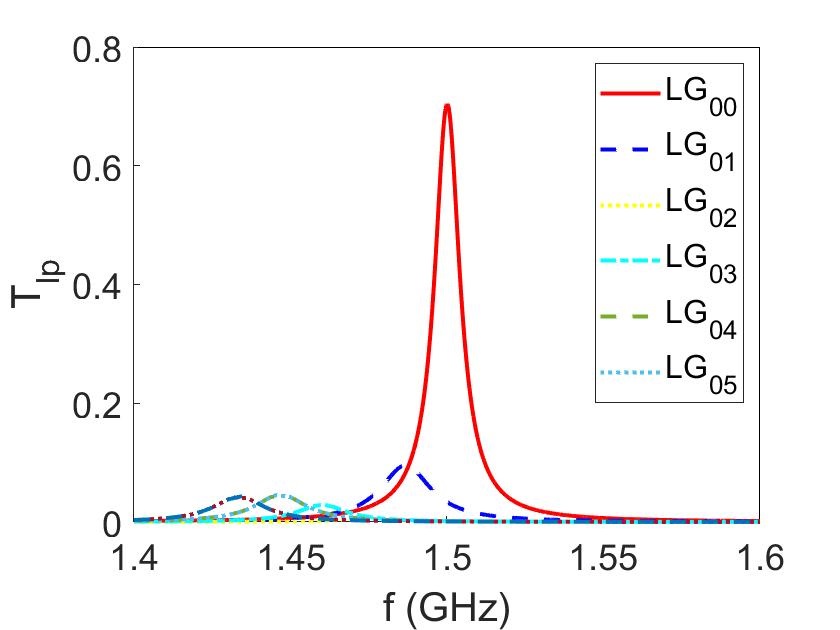}} 
\caption{The transmission of the optical drive through cavity mode $\text LG_{lp}$ is plotted against the frequency (f). The solid red curve is for the fundamental mode $\text LG_{00}$, the dashed blue curve is for the first higher order mode $\text LG_{01}$, the dotted pink curve is for the $\text LG_{02}$ mode and the dot-dashed cyan curve is for the $\text LG_{03}$ mode. Figures a) b) c) and d) demonstrate the cases where minimum beam waists are 5.7, 6.5, 7.3, and 9.3 $\mu$m, respectively.}
\label{Individual linewidth broadening}
\end{figure}

The linewidth broadening effect becomes more significant in the strong focusing regime, where beam waist at the mirror ($w_m$) is much larger than the beam waist at the focus, $w_0/w_m\ll1$, as shown in Fig.~\ref{Linewidth broadening}. In addition to Ref.~\cite{durak2014diffraction}, which examines the case of $w_0=5.7$ $\mu$m minimum beam waist value, $w_0=6.5 , 7.3$ and $9.3$ $\mu$m cases are examined to demonstrate the effect of different minimum beam waists on the cavity linewidth. Transmission coefficients for individual cavity modes $\text T_{lp}$, shown in Fig.~\ref{Individual linewidth broadening}, are closely spaced, being impossible to resolve altogether.

\begin{figure}
\subfloat[]{\includegraphics[width=1.8in]{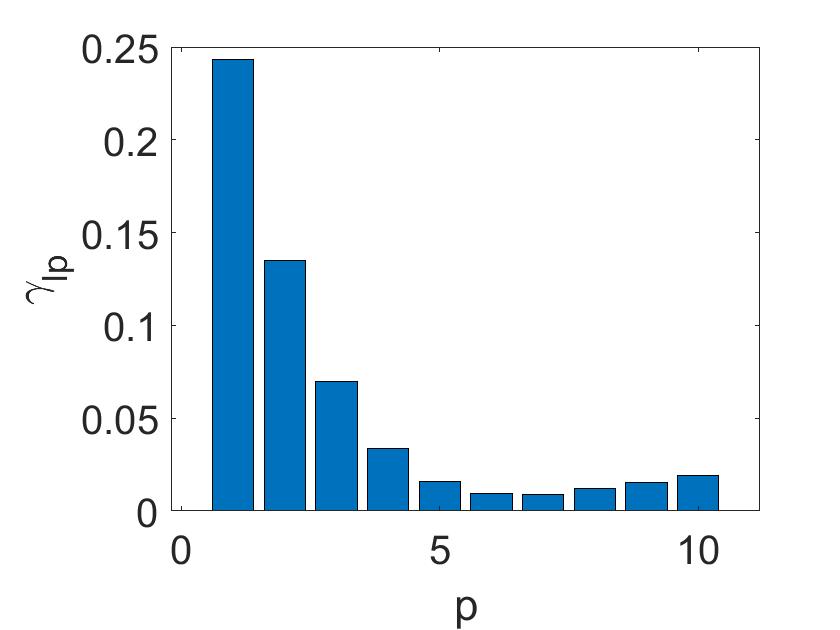}} 
\subfloat[]{\includegraphics[width=1.8in]{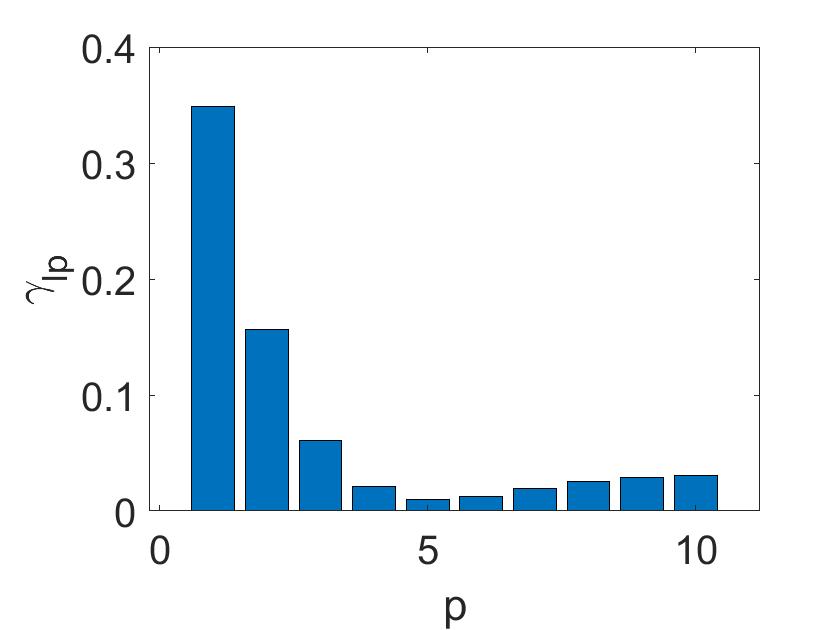}}\\
\subfloat[]{\includegraphics[width=1.8in]{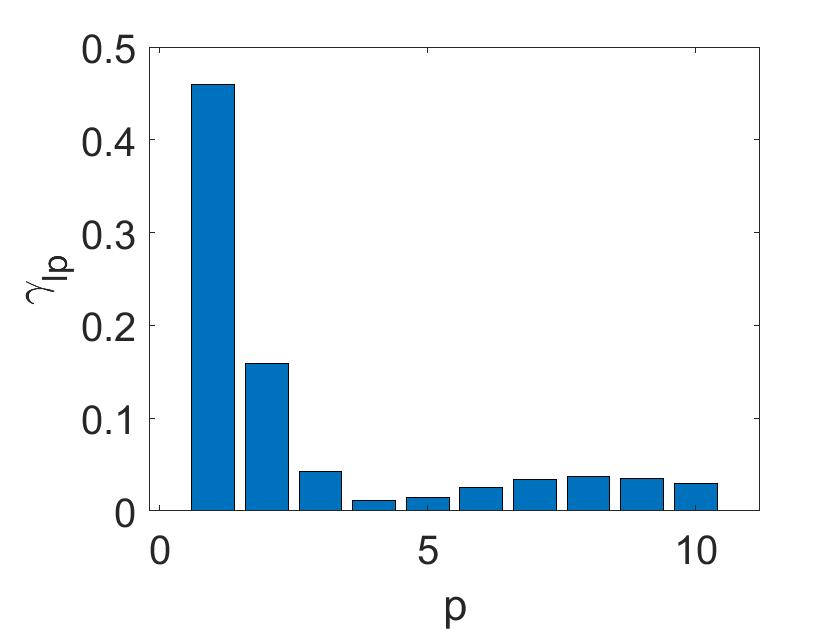}}
\subfloat[]{\includegraphics[width=1.8in]{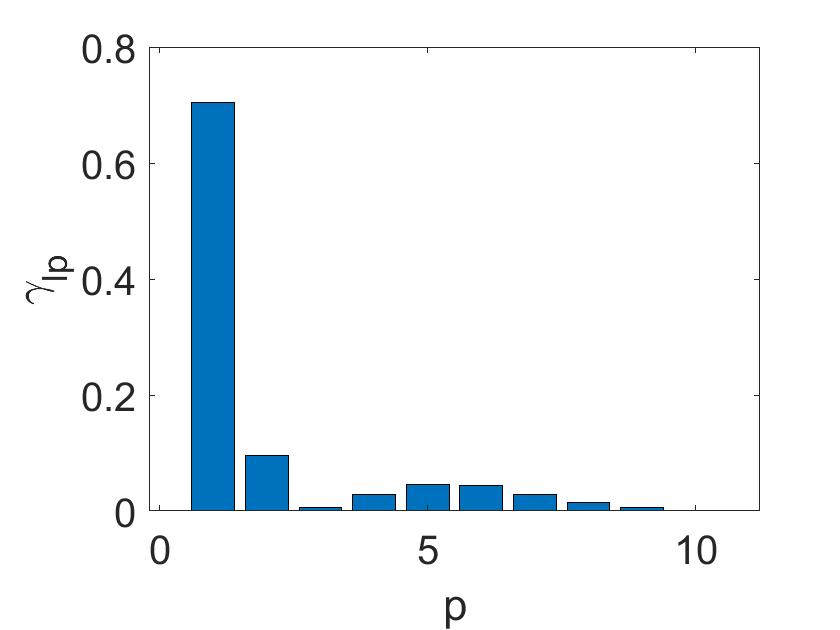}} 
\caption{The overlap function $\gamma_{lp}$ between the optical mode and cavity mode $\text LG_{lp}$. Figures a) b) c) and d) demonstrate the cases where minimum beam waists are 5.7, 6.5, 7.3, and 9.3 $\mu$m, respectively.}
\label{Coupling}
\end{figure}

\begin{figure}[th!]
\subfloat[]{\includegraphics[width=1.8in]{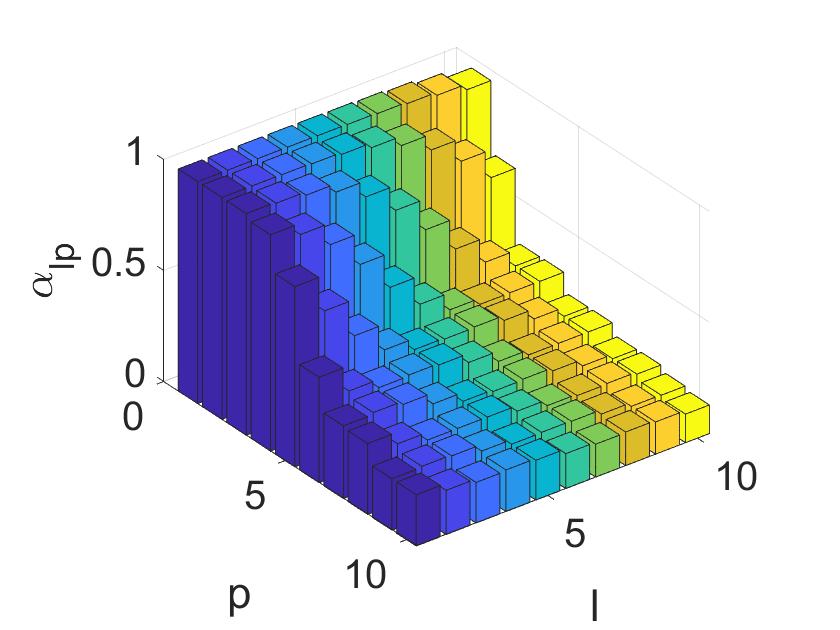}} 
\subfloat[]{\includegraphics[width=1.8in]{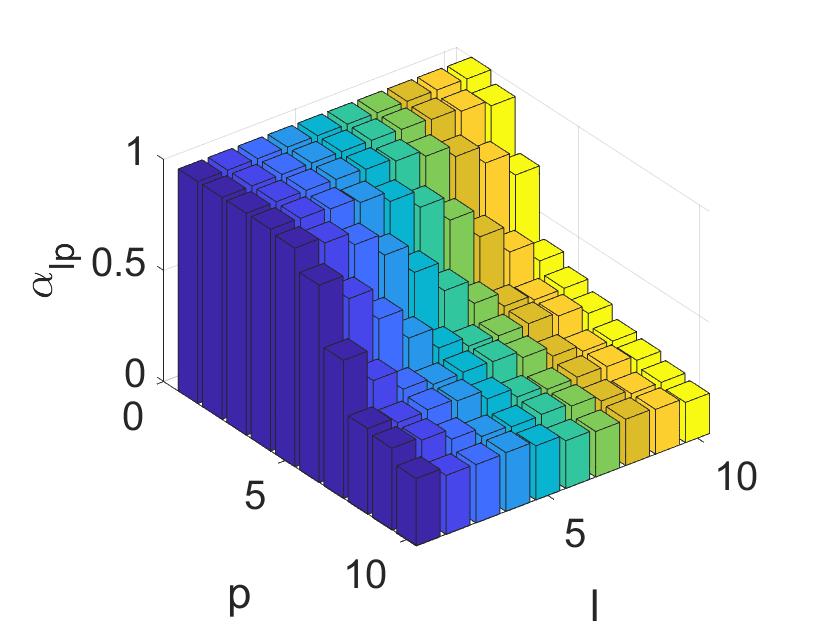}}\\
\subfloat[]{\includegraphics[width=1.8in]{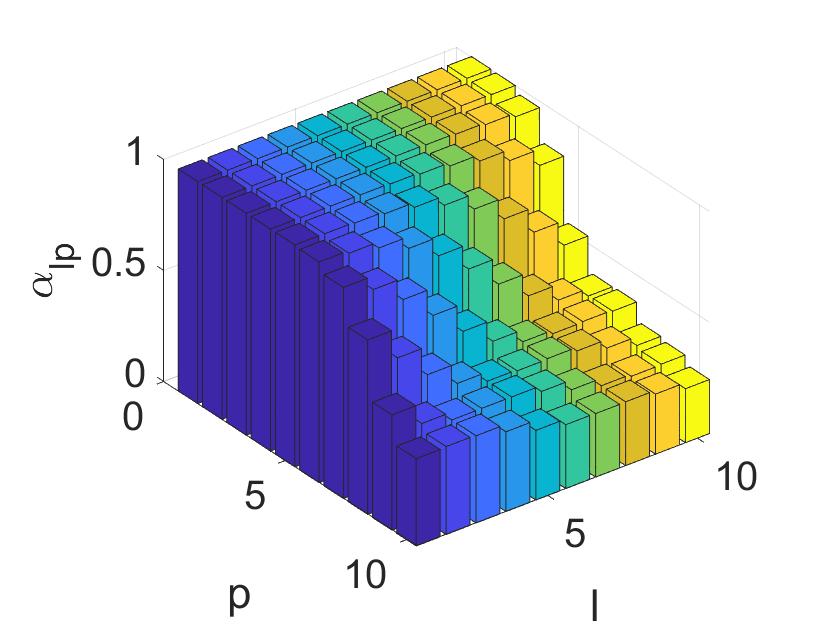}}
\subfloat[]{\includegraphics[width=1.8in]{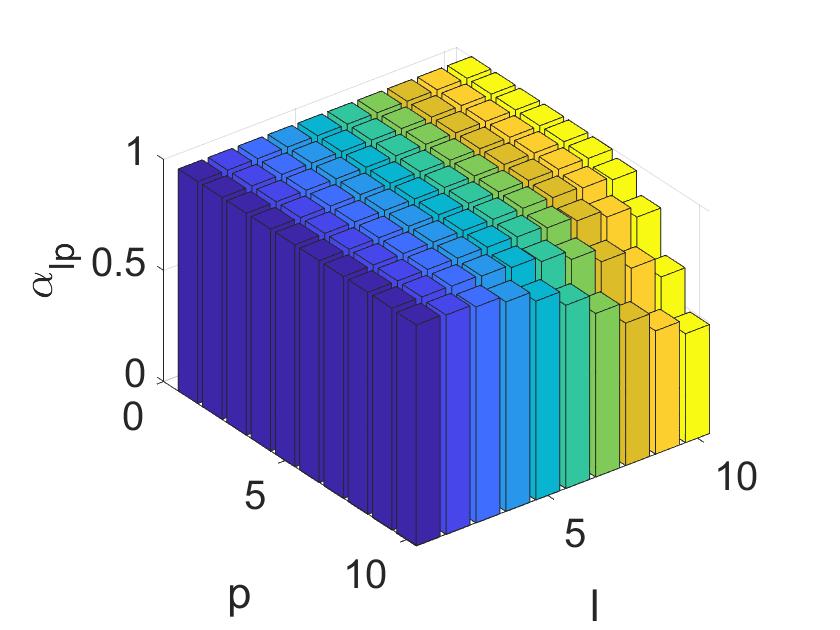}} 
\caption{Power left ($\alpha_{lp}$) in each cavity mode ($\text LG_{lp}$) after one round trip for different azimuthal and transverse mode indices $l$ and $p$ values. Figures a) b) c) and d) demonstrate the cases where minimum beam waists at the center of the cavity in the near-concentric regime are 5.7, 6.5, 7.3, and 9.3 $\mu$m, respectively.}
\label{Diffraction losses calculated}
\end{figure}

The coupling efficiencies ($\gamma_{lp} $) and the power left inside the cavity after one round trip ($ \alpha_{lp} $) for each higher-order modes are seen in Fig.~\ref{Coupling} and Fig.~\ref{Diffraction losses calculated}, respectively. Coupling efficiencies become minmum after $\text p=15$ which means higher-order mode excitations are remarkable in the first several cavity eigenmodes. As the focusing strength decreases, where sizes of the minimum beam waist and the beam waist at the mirror become comparable, coupling efficiency in the higher-order modes decrease. 

 Diffraction losses become dominant for the higher-order modes and it is the main reason for the power loss for those cavity eigenmodes. Due to the finite size of the mirrors, diffraction losses are inevitable and higher-order modes suffer from diffraction losses due to their larger size. The effect of the diffraction losses for the near-concentric cavity can be seen in Fig.~\ref{Diffraction losses calculated}. It is seen that if the minimum beam waist is larger, higher-order modes are less affected. The reason for this is, the beam waist at the mirror decreases with the increasing minimum beam waist which results in lower focusing power. With the decreasing beam waist at the mirror, one can observe that the effect of the diffraction losses becomes less important, and as a result, the broadening effect becomes manageable.

\begin{figure}
    \subfloat[]{\includegraphics[width=1.8in]{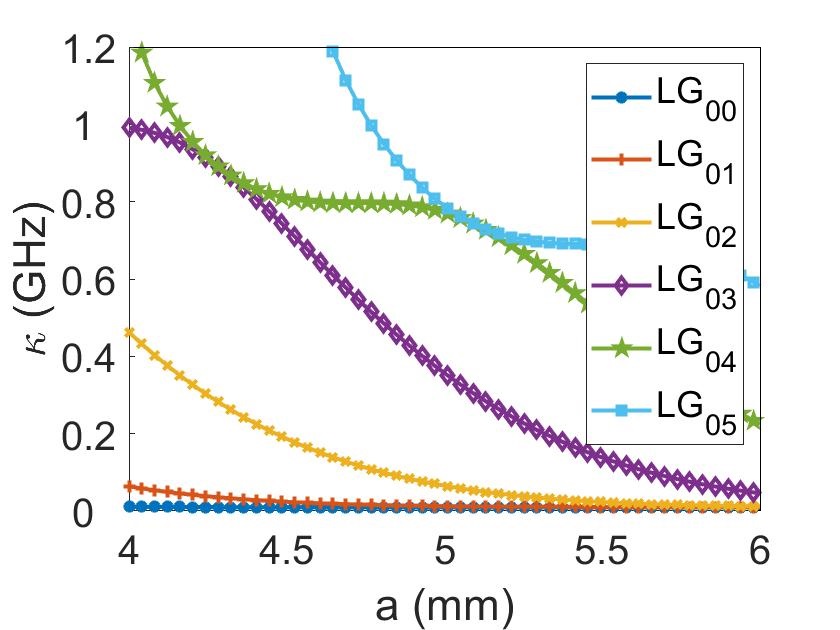}} 
    \subfloat[]{\includegraphics[width=1.8in]{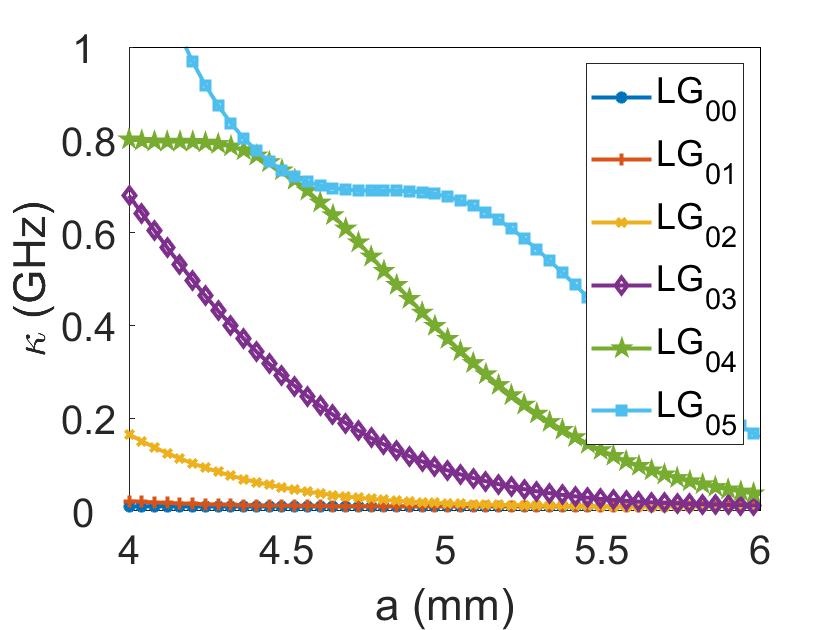}}\\
    \subfloat[]{\includegraphics[width=1.8in]{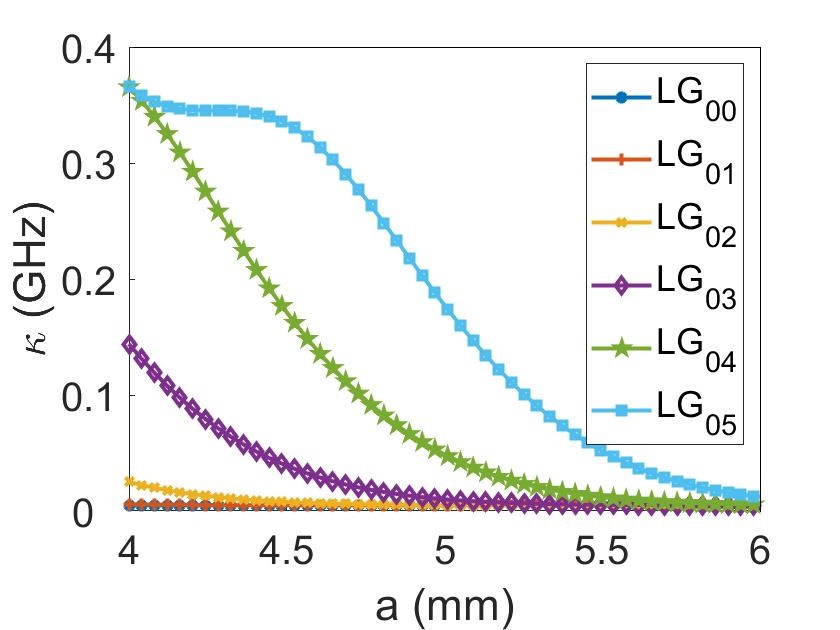}}
    \subfloat[]{\includegraphics[width=1.8in]{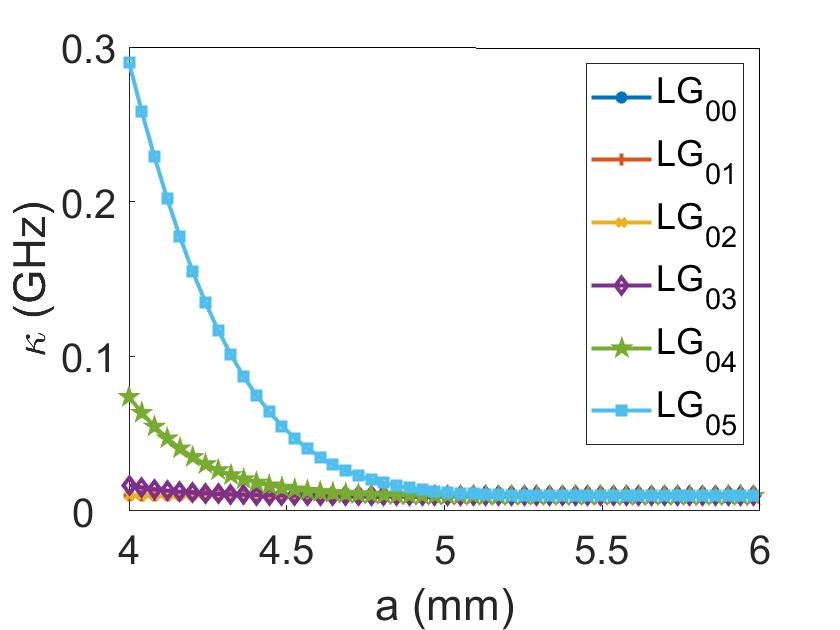}} 
    \caption{Dependence of the linewidth of the cavity modes $\text LG_{lp}$ on the aperture size ($a$). The circle marker blue curve is for the fundamental mode $\text LG_{00}$, the rectangular marker orange curve is for the first higher order mode $\text LG_{01}$, the cross marker yellow curve is for the $\text LG_{02}$ mode and the diamond marker purple curve is for the $\text LG_{03}$ mode. $\kappa$ is the linewidth in the units of GHz and $a$ is the aperture size in the units of mm. }
    \label{Aperture size vs linewidth}
\end{figure}

\begin{figure}
    {\includegraphics[width=2.5in]{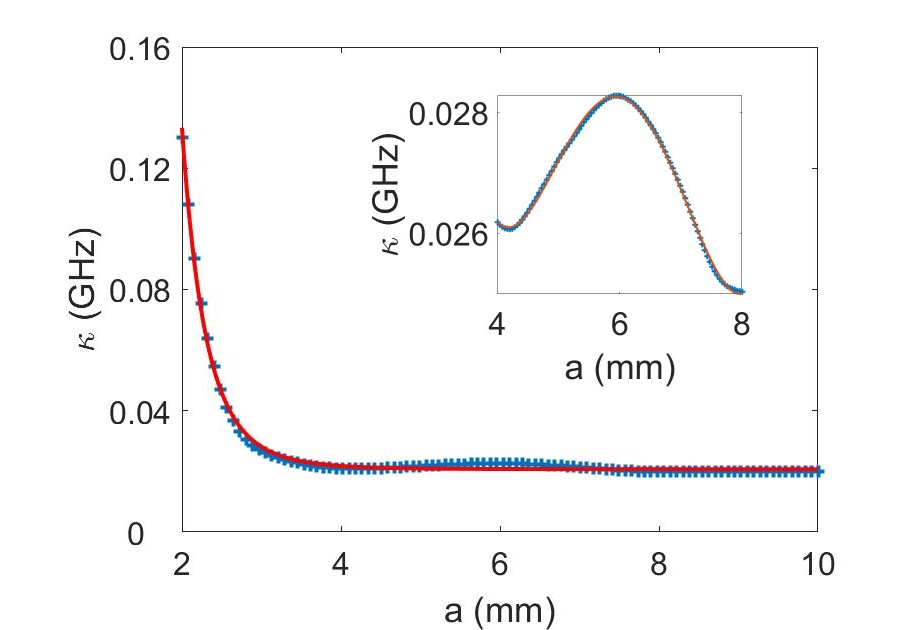}} 
    \caption{Dependence of the cavity linewidth on the aperture size. The left figure shows the behavior for the aperture sizes between 2 to 10 mm. On the right, a zoomed-in version can be seen which shows the behavior for the aperture sizes between 4 and 8 mm. $\kappa$ is the linewidth in the units of GHz and $a$ is the aperture size in the units of mm.}
    \label{Aperture size vs overall cavity linewidth}
\end{figure}

The effect of the mirror apertures on the linewidth of each higher-order mode can be seen in Fig.~\ref{Aperture size vs linewidth}. As aperture size increases, linewidth for each cavity eigenmode decreases and this behavior does not change for different focusing strengths. Changes in the focusing strength can easily be seen in the different minimum beam waist. When  each higher-order mode effect is taken into account, overall cavity linewidth behavior can be calculated. The overall behavior of the cavity when aperture size is increasing is seen in Fig.~\ref{Aperture size vs overall cavity linewidth}. Aperture size ranges of 2-10 mm and 4-8 mm are demonstrated in Fig.~\ref{Aperture size vs overall cavity linewidth}.

\begin{figure}
\subfloat[]{\includegraphics[width=1.8in]{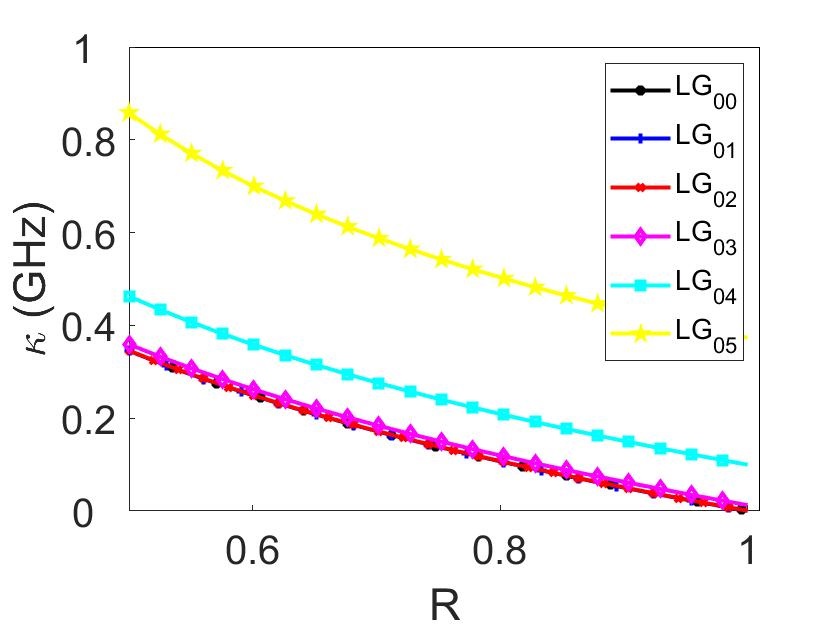}} 
\subfloat[]{\includegraphics[width=1.8in]{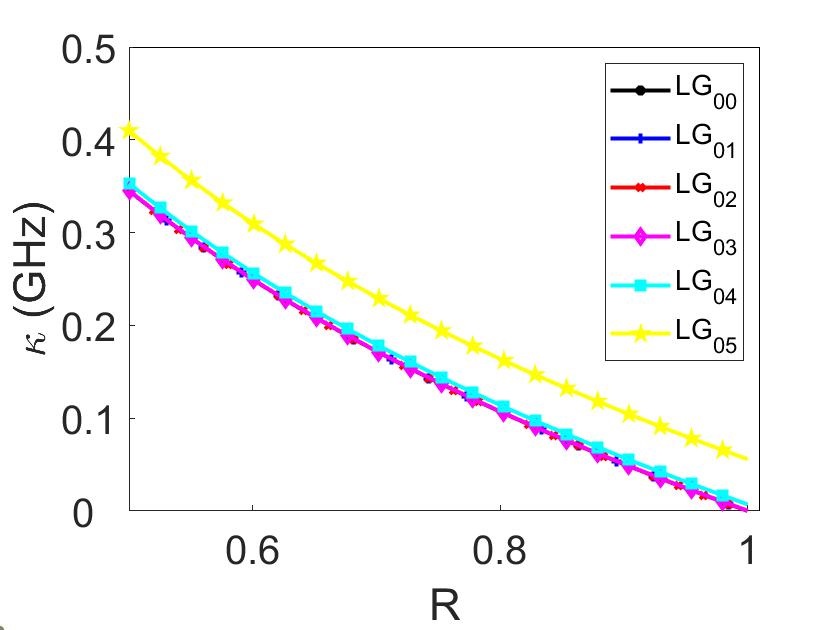}}\\
\subfloat[]{\includegraphics[width=1.8in]{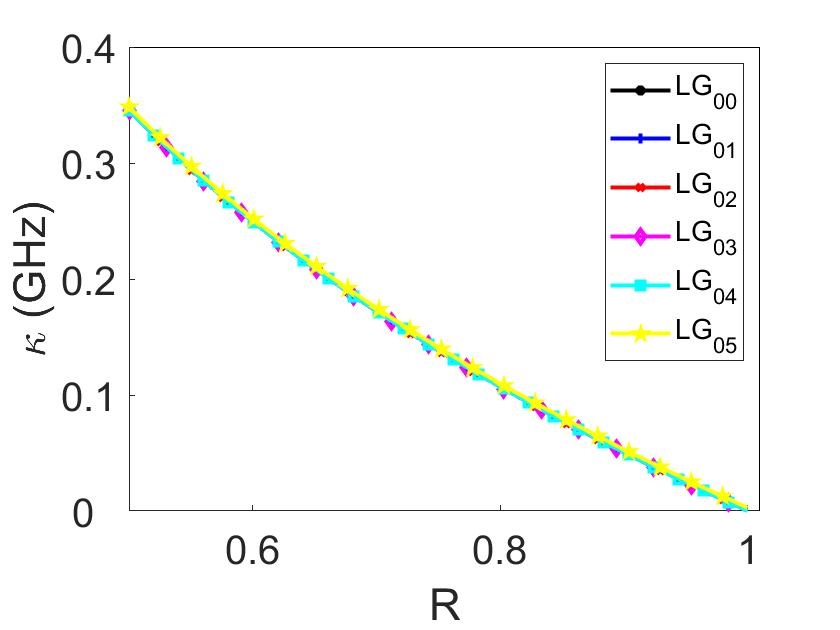}}
\subfloat[]{\includegraphics[width=1.8in]{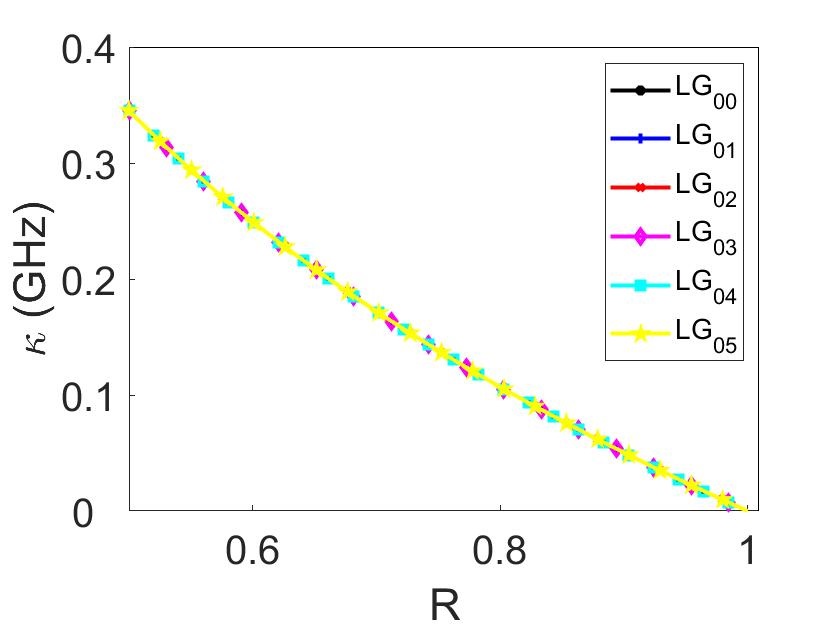}} 
\caption{Linewidth vs reflectivity. Figures a) b) c) and d) demonstrate the cases where minimum beam waists are 5.7, 6.5, 7.3, and 9.3 $\mu$m, respectively. The circle marker black curve is for the fundamental mode $\text LG_{00}$, the rectangular marker blue curve is for the first higher order mode $\text LG_{01}$, the cross marker red curve is for the $\text LG_{02}$ mode and the diamond marker purple curve is for the $\text LG_{03}$ mode. $\kappa$ is the linewidth in the units of GHz and $R$ is the reflectivity given in the range of 0-1.}
\label{Linewidth vs reflectivity}
\end{figure}

\begin{figure}
    \centering
    \includegraphics[width=2.5in]{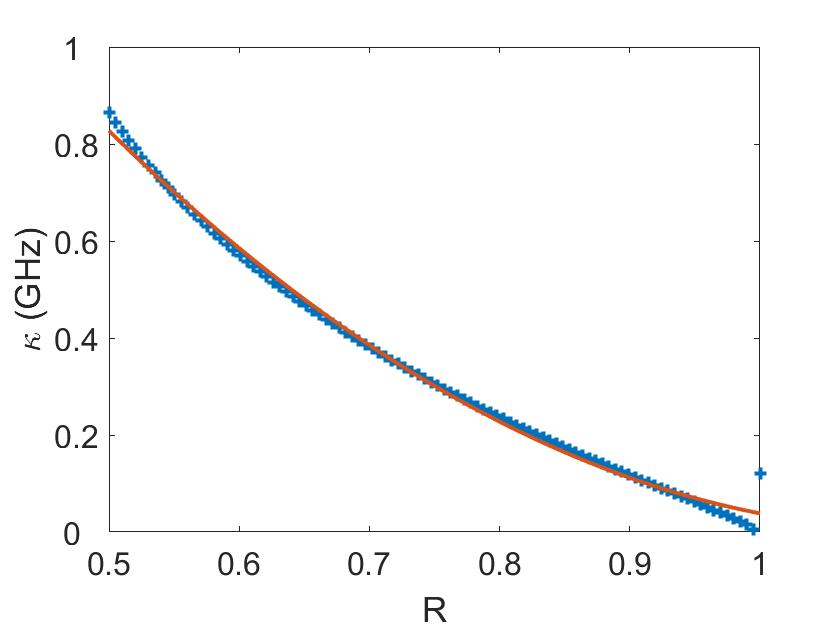}
    \caption{Overall cavity linewidth changing with increasing effective mirror reflectivity. Data points are shown with the blue daggers whereas the fit is shown with the red solid curve. $\kappa$ is the linewidth in the units of GHz and $R$ is the reflectivity given in the range of 0-1.}
    \label{Reflectivity vs overall linewidth}
\end{figure}

One of the other most important factors that contribute to the linewidth of the cavity is the reflectivity of the mirrors which is seen in Fig.~\ref{Linewidth vs reflectivity}. While calculating the linewidth, not only the reflectivity of the mirrors are taken into account but also reflectivity affected by the diffraction losses are calculated which is $R_\text{eff}=\rho R$. It can be seen that the behavior is the same for the first few cavity eigenmodes in Fig.~\ref{Linewidth vs reflectivity}. The first few cavity eigenmodes have comparable sizes, while higher-order modes have larger sizes and suffering from diffraction losses. With diffraction losses taken into account, effective reflectivity changes for higher-order cavity eigenmodes, resulting in broadened linewidth for these modes. Overall cavity linewidth changes with the mirror reflectivity can be seen in Fig.~\ref{Reflectivity vs overall linewidth}. As expected, linewidth decreases with increasing reflectivity, whose behavior can be described by second-order polynomial fit. 

\begin{figure}
    \centering
    \includegraphics[width=2.5in]{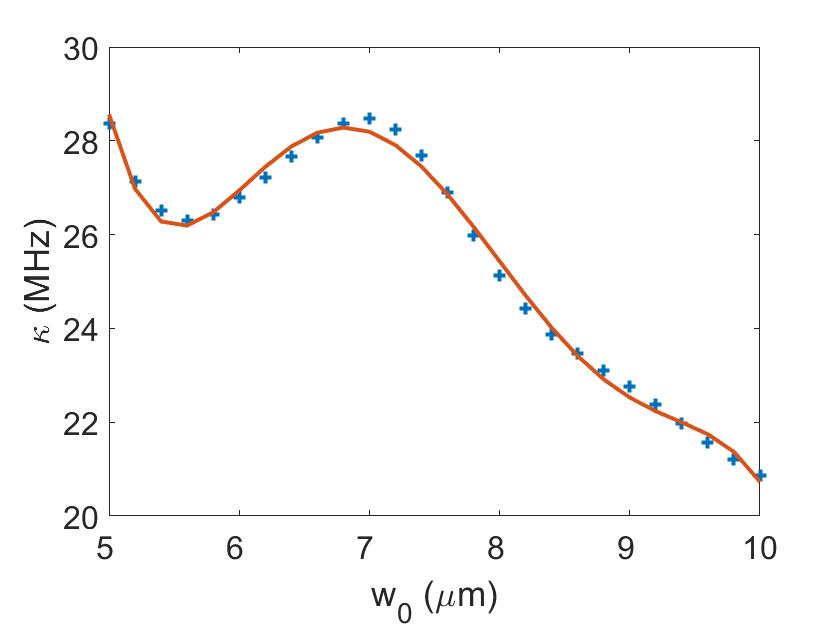}
    \caption{The figure shows how overall cavity linewidth changes with the increasing minimum beam waist. For the calculation, half cavity length and beam waist at the mirror are calculated separately for each corresponding minimum beam waist value and fitted accordingly. $\kappa$ is the linewidth in the units of MHz and $w_0$ is the minimum beam waist in the units of $\mu m$. }
    \label{Minimum beam waist vs linewidth}
\end{figure}

Focusing strength is inversely proportional to the minimum beam waist of the cavity. A bigger minimum beam waist means lower focusing strength which results in a smaller beam waist at the mirror. When the beam waist at the mirror is small, the losses due to the finite aperture size of the mirrors does not affect each higher-order mode differently and the linewidth broadening effect is negligible. Another reason is the validity of the paraxial approximation in that regime. Paraxial approximation holds when $w_0 \gg \lambda$, which means low focusing strength. The linewidth broadening effect can be seen for different minimum beam waist values which mean different focusing strengths in Fig.~\ref{Minimum beam waist vs linewidth}. As the focusing strength decreases, linewidth also decreases due to the decreasing higher-order mode excitation. For different minimum beam waist values, half cavity lengths and beam waist at the mirror values are calculated and overall cavity calculations are done according to those values in Fig.~\ref{Minimum beam waist vs linewidth}.

\section{\label{sec:conclusion} Conclusion}

In summary, we presented a numerical and experimental investigation of the effects of the higher-order modes in an optical cavity in the strong focusing regime. Different than earlier works ~\cite{durak2014diffraction,kleckner2010diffraction}, our results include the effects of the different parameters like reflectivities, numerical apertures, and beam waist sizes for each higher-order mode. Effective and complete treatment of the near-concentric cavities can be done with the help of the mode decomposition analysis applied for those features. The excitation of higher order modes, due to diffraction loss or any other imperfect mode-matching mechanism, can result in the broadening of the cavity transmission peak effectively. This effect is difficult to distinguish experimentally and the proposed method allows the realization of the effects of the higher order modes. The optical mode decomposition into the cavity eigenmodes provides means to simulate the intensity and transmission spectrum distributions of any cavity. The assumptions of the model seem to be justified by a comparison of experimental results with the simulated ones in the near-concentric regime. The choice of near-concentric is due to the fact that the effect of linewidth broadening is more prominent in this regime. The presented model is expected to guide the experimentalists in the preparation of cavities for avoiding misinterpretations of the observed cavity outputs in the experiment of interest. Our results can illuminate the way of designing optical cavities for strong coupling cavity QED applications, demanded by modern quantum technologies. 

\begin{acknowledgments}
This work was supported by the Scientific and Technological Research Council of Turkey (TUBITAK) with project number 119F200.
\end{acknowledgments}

\bibliography{apssamp}

\end{document}